\documentclass[oldversion]{aa}
\usepackage{graphicx}
\usepackage{txfonts}
\begin{document}
\title{A Suprime-Cam study of the stellar population of Ursa Major I dwarf spheroidal galaxy\thanks{Based on data collected at Subaru Telescope, which is operated by the Nationa Astronomical Observatory of Japan.}}

%   \subtitle{I. Overviewing the $\kappa$-mechanism}

\author{S. Okamoto\inst{1,2}, 
            N. Arimoto\inst{2,3}, 
            Y. Yamada\inst{2}
            \and
            M. Onodera\inst{4}
            }

\titlerunning{The stellar population of Ursa Major I dSph}
\authorrunning{S. Okamoto et al.}
\institute{Department of Astronomy, University of Tokyo, Hongo, Bunkyo-ku, Tokyo 113-0033, Japan
         \and
             National Astronomical Observatory of Japan, Osawa 2-21-1, Mitaka, Tokyo 181-8588, Japan
         \and
            The Graduate University for Advanced Studies, Osawa 2-21-1, Mitaka, Tokyo 181-8588, Japan
         \and
            Institute of Earth, Atmosphere and Astronomy, BK21 Yonsei University, 134 Sinchon-dong, Seodaemun-gu, Seoul, 120-749, Republic of Korea
}
\offprints{S. Okamoto,\\
    \email{sakurako.okamoto@nao.ac.jp}}

\date{Accepted : 4 April 2008}

% \abstract{}{}{}{}{} 
% 5 {} token are mandatory

\abstract{We present deep and wide V, I CCD photometry of Ursa Major I (UMa I) dwarf spheroidal galaxy (dSph) in Local Group.  The images of the galaxy were taken by Subaru/Suprime-Cam wide field camera,  covering a field of 34\arcmin $\times$ 27\arcmin~located at the centre of the galaxy.  Colour-magnitude diagram (CMD) of the UMa I dSph shows a steep and narrow red giant branch (RGB), blue and red horizontal branch (HB), and main sequence (MS) stars.  A well-defined main sequence turn-off (MSTO) is found to be located at V$_{0,MSTO}\sim$23.5 mag.  The distance modulus is derived as $(m-M)_0=19.93\pm0.1$ (corresponding to a distance D$=96.8\pm4$ kpc) from the V-band magnitude of the horizontal branch (V$_{0,HB}=20.45\pm0.02$).  The mean metallicity of the RGB stars is estimated by the V$-$I colour to [Fe/H]$\sim-2.0$.  The turn-off age estimated by overlaying the theoretical isochrones reveals that most of stars in the UMa I dSph are formed at very early epoch ($\sim12$Gyrs ago).  The isopleth map of stellar number density of the UMa I dSph, based upon the resolved star counts of MS, RGB, HB stars as well as blue stragglers (BS), shows that the morphology of the UMa I dSph is quite irregular and distorted, suggesting that the galaxy is in a process of disruption.  
The very old and metal-poor nature of the stellar population implies that the star formation history of this newly discoverd faint dSph may have been different from other well-known `classical' dSphs, which show significant stellar population of intermediate age.
The stellar population of the UMa I dSph closely resembles that of Galactic old metal-poor globular cluster, but its size is typical of Galactic dSphs (r$_{e}$=188 [pc], r$_{1/2}$=300 [pc]), and the shape of its spatial density contours suggests that it is undergoing tidal disruption. 
These characteristics of stellar population and spatial distribution of the faint galaxies help us to know how they formed and evolved, and give a hint to the nature of the building blocks of hierarchical galaxy formation. 
} 


\maketitle
%
%________________________________________________________________

\section{Introduction}

The origin of Galactic dwarf spheroidal (dSph) galaxies should closely be related to the formation and evolution history of the Milky Way.  Modern cosmological models based on the Cold Dark Matter paradigm demonstrate the importance of hierarchical structure formation in all scales.  Galaxies like the Milky Way and M 31 form as a part of a local overdensity in the primordial matter distribution via the agglomeration of numerous smaller building blocks which independently could develop into dwarf galaxies if they could have been escaped from any take over by bigger ones.  The relatively gas-rich dwarf irregulars still exhibit on-going star formation, while the dwarf spheroidals, being devoid of significant amounts of gas and dust, are now quiescent and are therefore, in principle, much simpler systems to study.  The proximity of Galactic dSphs offers a unique opportunity for investigating galaxy formation and evolution in unprecedented detail by studying the photometric and spectroscopic properties of stellar populations. 

Analyses of resolvable stellar distribution in the Sloan Digital Sky Survey (SDSS) data archive have led to recent discoveries of numerous very faint Galactic satellites (Willman et al. \cite{wil05}; Belokurov et al. \cite{bel06}; Zucker et al. \cite{zuc06a}, \cite{zuc06b}; Belokurov et al. \cite{bel07}; Irwin et al. \cite{irw07}; Walsh et al. \cite{wal07}).  The Ursa Major I (UMa I) dwarf spheroidal galaxy locating about 100 kpc from the Milky Way is the first dSph discovered by SDSS (Willman et al. \cite{wil05}).  The absolute magnitude of the UMa I dSph is $M_V\sim-6.75$,  which is about 8 times less luminous than the faintest ``classical" Galactic dSphs, such as the Draco, Ursa Minor, and Sextans dSph.  Measuring velocities of the UMa I stars, Kleyna et al. (\cite{kle05}) and Martin et al. (\cite{mar07}) suggested that the UMa I dSph might be the most dark matter dominated object in the Universe. 

The properties of newly discovered faint dSphs are poorly understood.  Because of their  faint luminosity and the size in the sky, the stellar ages, metallicities and structures of such dSphs are still unclear from previous limiting observations.  The super-wide field of view and enough depth of the Subaru/Suprime-Cam offers a unique opportunity for successfully and efficiently addressing the photometric aspect of this research.

This paper is organised as follows: In Section 2, we describe the observation and data analysis procedure. In Section 3, we present colour-magnitude diagrams (CMDs) of the UMa I dSph,  and discuss the stellar population of the UMa I dSph.  In Section 4, we estimate the age and metallicity of the dominant stellar population.  In Section 5, we analyse projected density profile based upon the resolved star counts. Then, we discuss the structure and evolution of the UMa I dSph and summarise this study in Section 6.

%__________________________________________________________________

\section{Observation and Data Reduction}

\begin{figure}
  \resizebox{\hsize}{!}{\includegraphics{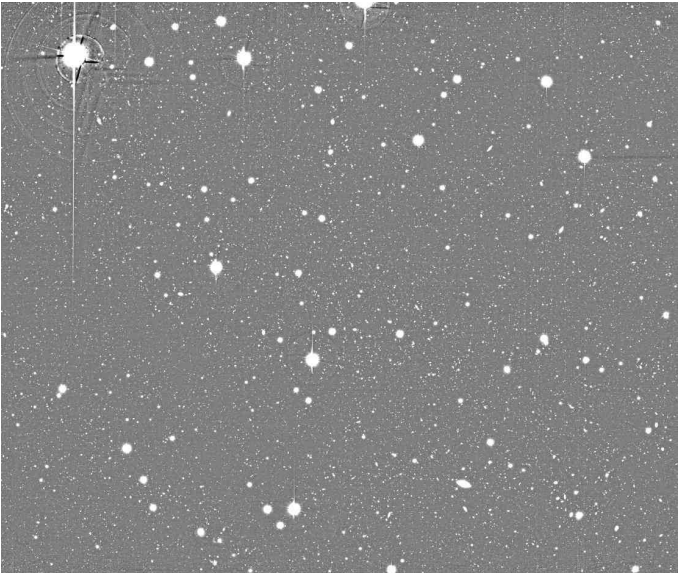}}
  \caption{The V-band Suprime-Cam image of Ursa Major I dSph covering a field of 34\arcmin $\times$ 27\arcmin.  North is to the top and east is to the left.}
  \label{fig:uma fov}
\end{figure}

We obtained deep V, I CCD images of the central part of the UMa I dSph with the Suprime-Cam (Miyazaki et al. \cite{miy02}) on the Subaru Telescope during nights of 2005 December 30 to 2006 January 2 (PI. N. Arimoto).  The Suprime-Cam consists of a 5 $\times$ 2 arrays of 2048 $\times$ 4096 CCD detectors and provides a field-of-view of 34\arcmin $\times$ 27\arcmin with a pixel scale of 0.202\arcsec. 
The nights of the observing run were photometric and the seeing ranged from 0.5\arcsec to 0.8\arcsec.  To avoid saturation of bright stars, we took long and short exposure images with Johnson V (10 $\times$ 50s and 6 $\times$ 10s) and Cousins I (10 $\times$ 110s and 3 $\times$ 30s) filters.

Data were processed using a pipeline software SDFRED dedicated to the Suprime-Cam (Yagi et al. \cite{yag02}; Ouchi et al. \cite{ouc04}).  Each image was bias-subtracted and trimmed, flat-fielded, distortion and atmospheric dispersion corrected, sky-subtracted, and combined in usual manner. Astrometric calibration of each passband was based on a general zenithal polynomial projection derived from astrometoric standard star selected from online USNO catalog\footnote{http://ftp.nofs.navy.mil/data/fchpix/}.  Figure \ref{fig:uma fov} shows the V-band combined image of the UMa I dSph.  Because of its faint luminosity, we cannot see the shape of the galaxy directly.  

For these processed images, the DAOPHOT in IRAF package was used to obtain the point-spread-function (PSF) photometry of the resolved stars (Stetson \cite{ste87}).  In order to separate the point sources from the extended ones and noise-like objects, we used DAOPHOT parameter $\chi^2$ and SHARP with the following cuts: $\chi^2$ $<$ 4 and $-0.4$ $<$ SHARP $<$ 0.4.  The position of detected stellar objects in each band was cross-correlated (within 1\arcsec) in order to make catalogues of long and short exposure images.  Then, combining bright stellar objects (V $<$ 21) detected in short exposure image and faint stellar objects (V $>$ 21) detected in long exposure image, we obtained a candidate list of about 2800 stars.  

Instrumental magnitudes of sources in the CCD images were calibrated to the standard Johnson-Cousins photometric system using the photometric standard stars of Landolt (\cite{lan92}), observed during the same nights.  To confirm our photometric calibration, we checked the consistency of the magnitude of stars commonly detected in both long and short exposure images (typically 20-22 mag stars).  The difference between the magnitude of the long and short exposure is $\Delta M_{short-long} < 0.02$. The average extinction, E(B$-$V)$=0.019\pm0.005$ (corresponding to A$_V=0.06\pm0.01$, A$_I=0.04\pm0.01$\footnote{The assumed extinction law is the R$_v$=3.1 (Cardelli et al. \cite{car98}) and the relation of A$_I$/A$_V$=0.594 (Schlegel et al. \cite{sch98}).}) 
in the direction of the UMa I dSph is taken by Schlegel et al. (\cite{sch98}).  

\begin{figure}
  \resizebox{\hsize}{!}{\includegraphics[angle=-90]{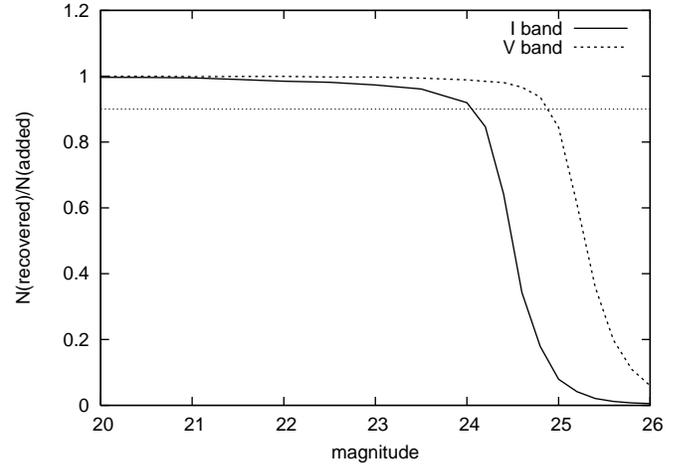}}
  \caption{Completeness of I (solid line) and V (doted line) photometry. The horizontal doted line represents the 90$\%$ level.}
  \label{fig:completeness}
\end{figure}

We derived completeness and photometric errors of our V, I photometry using the artificial-star test with the ADDSTAR routine in DAOPHOT.  We added 7000 artificial stars  to each CCD image in each 0.5 magnitude interval from 18 mag to 24 mag and in every 0.2 magnitude interval from 24 mag to 26 mag.  We processed the resulting images containing artificial stars in the same way as for the original images.  The detection ratio of the test, N(recovered)/N(added), is plotted in figure \ref{fig:completeness}, which shows that our photmetry is 90$\%$ complete at V=24.9 mag and I=24.1 mag, respectively, 
independently on position in the galaxy. 

The mean errors in the photometry are based on the difference between the input magnitude and the output magnitude of the artificial stars.  These errors are ploted in  figure \ref{fig:cmd} in the next section.  

%__________________________________________________________________

\section{Stellar Population of the UMa I dSph}

\begin{figure*}
\centering
\resizebox{\hsize}{!}{
  \includegraphics{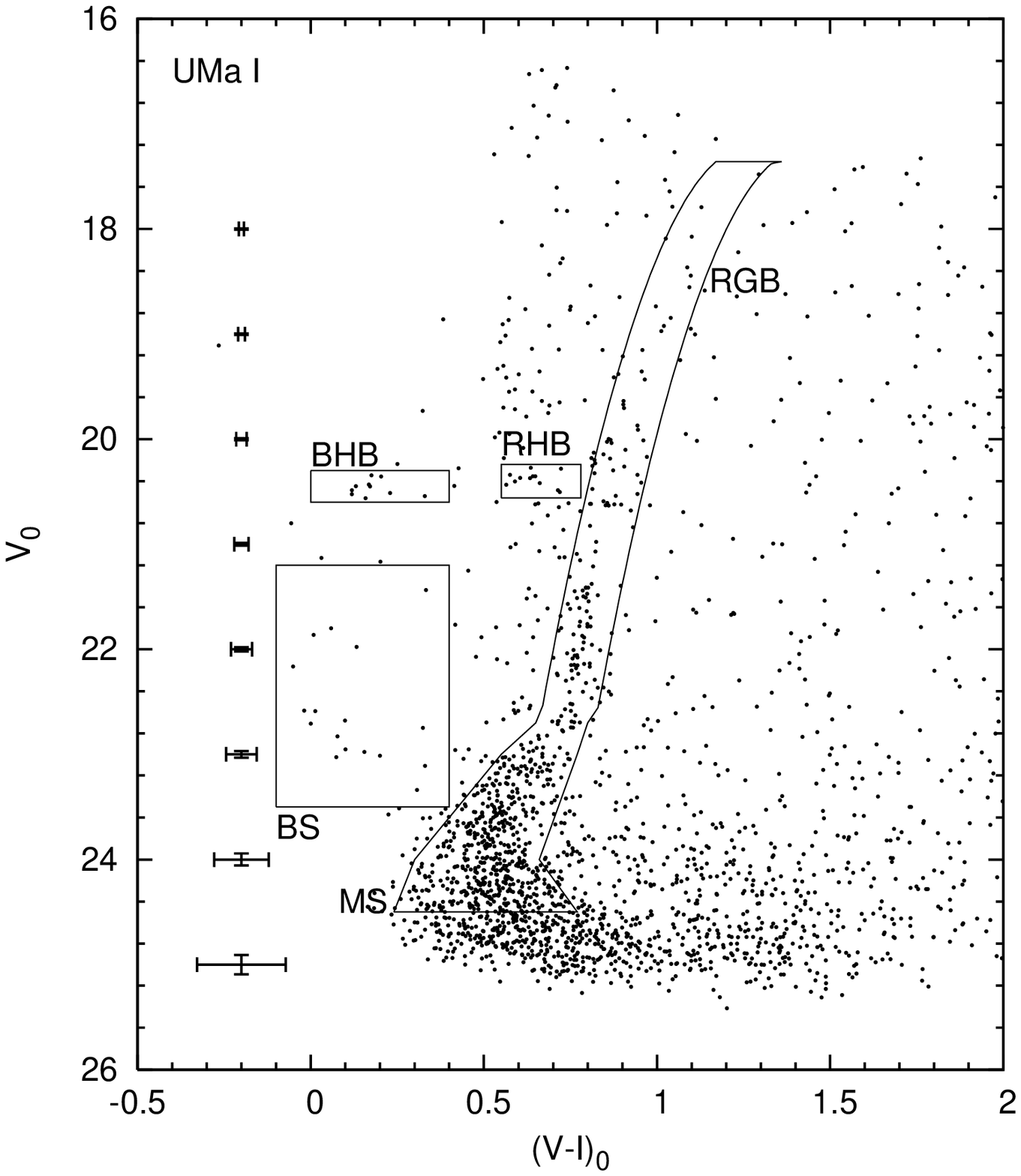}
  \includegraphics{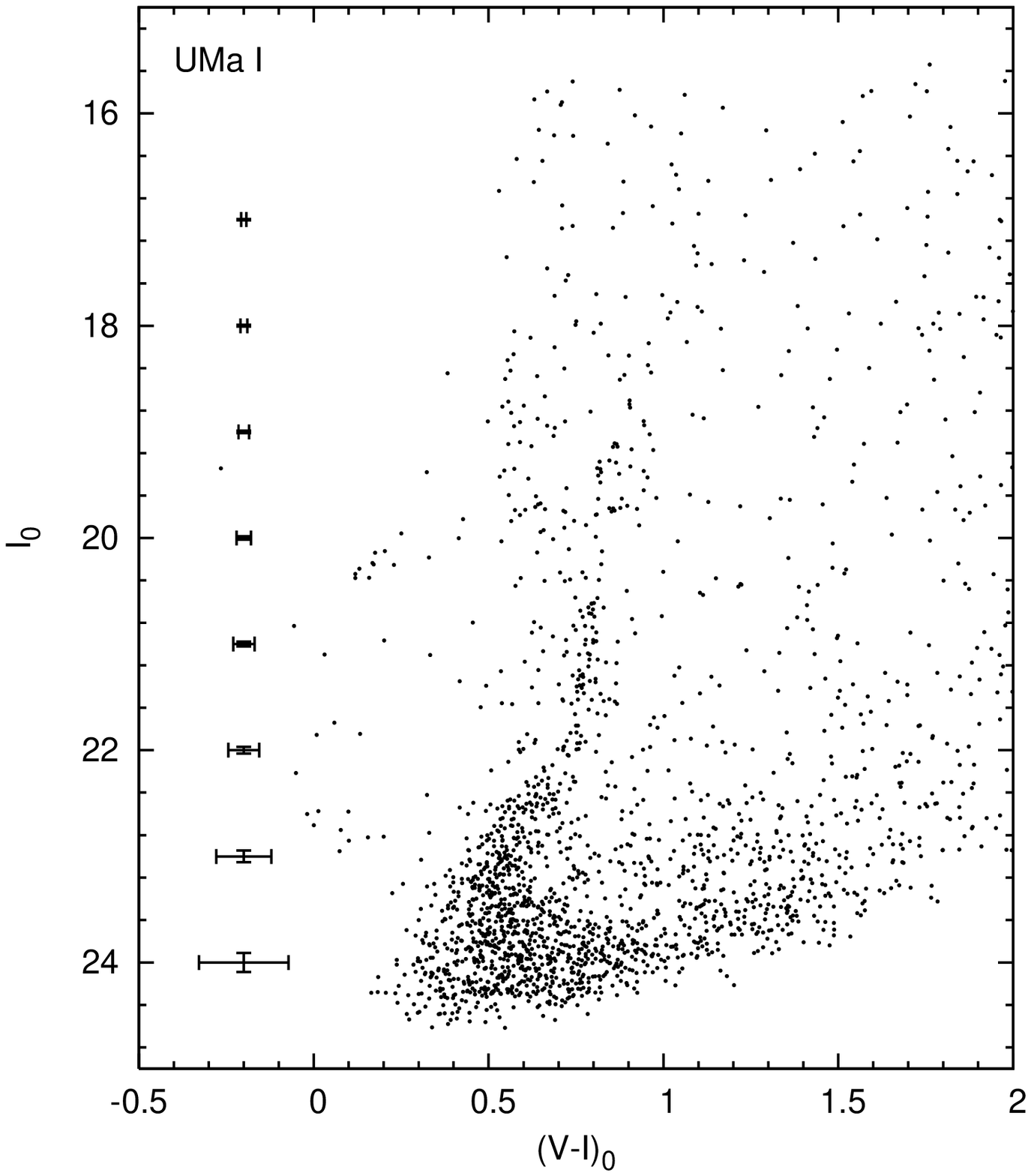}
}
\caption{Colour-magnitude diagrams of the UMa I dSph.  Error bars show the photometric error at each magnitude level based on the artificial star test. The boundaries marked in the top-left panel are used to select stars belonging to the main sequence, giant branch and horizontal branch and blue strugglers of the UMa I dSph. }
\label{fig:cmd}
\end{figure*}

Figure \ref{fig:cmd} show the de-reddened CMDs of the star-like objects found in the UMa I field.  Error bars in both figures show the photometric errors at each magnitude level based on the artificial star test.  These data are the first to reach $\sim$1 mag below the main sequence turn-off (MSTO), and one can easily see main sequence (MS) stars of the UMa I dSph.  
The major features of CMDs are as follows; 
(i) The red giant branch (RGB) is steep and narrow.  The tip of RGB is buried in the foreground stars.  The intrinsic width of RGB indicate a homogeneous chemical composition of the UMa I dSph.  
(ii) The horizontal branch (HB) stars are seen at V$_{0}$ $\sim$ 20.45 mag.  
(iii) Well-defined MS stars can be traced below V$_{0}$ $\sim$ 23 mag, with the MSTO located at V$_{0}$ $\sim$ 23.5 mag and (V$-$I)$_{0}$ $\sim$ 0.5 mag. 
(iv) There are a few blue straggler (BS) candidates. Although it has not been addressed whether these stars are young MS stars or BSs, the paucity of BS candidates in UMa I dSph suggests that the galaxy had no multiple episodes of star formation.

The CMDs are similar to those of metal-poor Galactic globular clusters, implying that the UMa I dSph has very old and metal-poor stellar population (see Section 4).  

\begin{figure*}
\centering
\resizebox{\hsize}{!}{
  \includegraphics{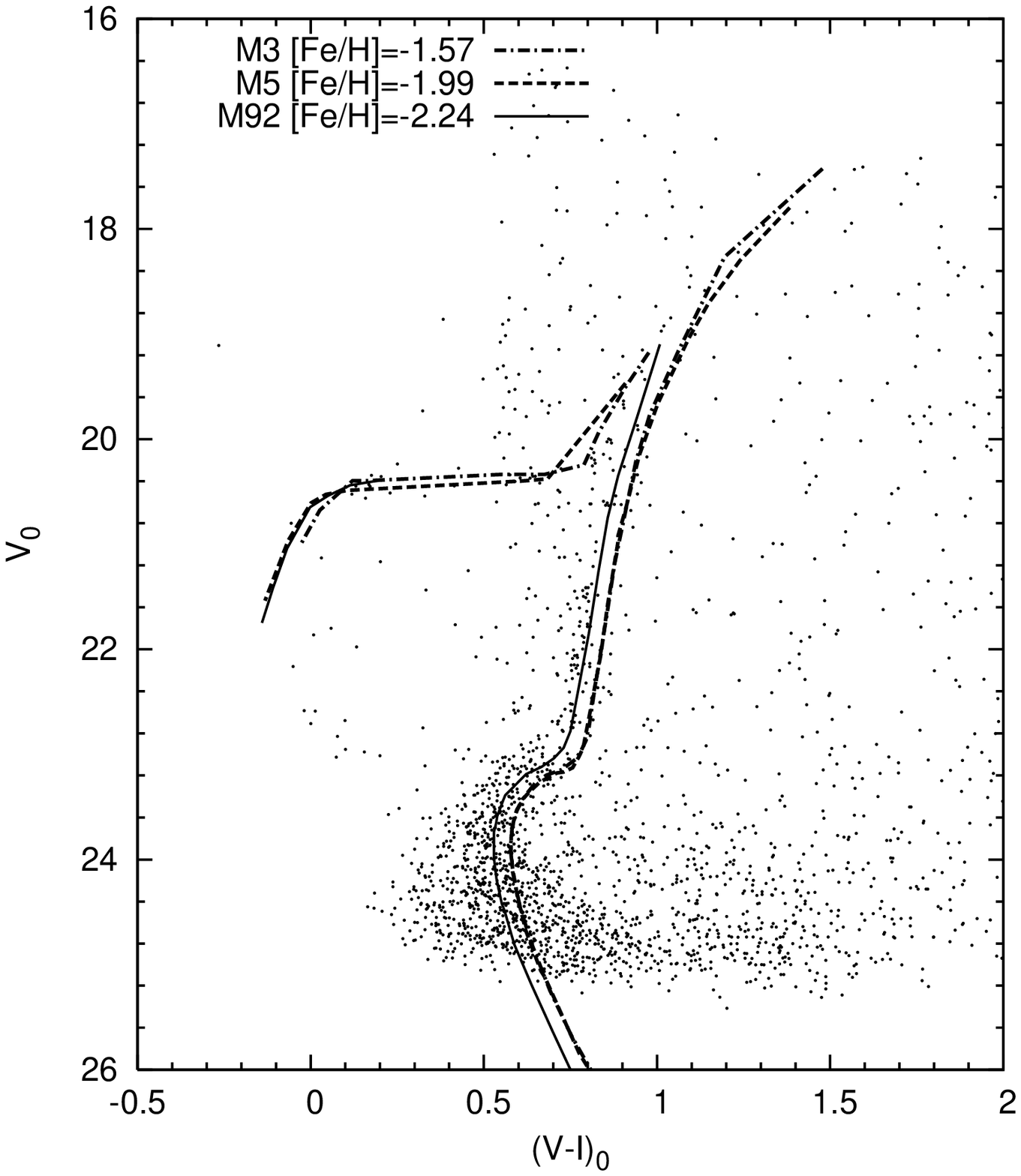}
  \includegraphics{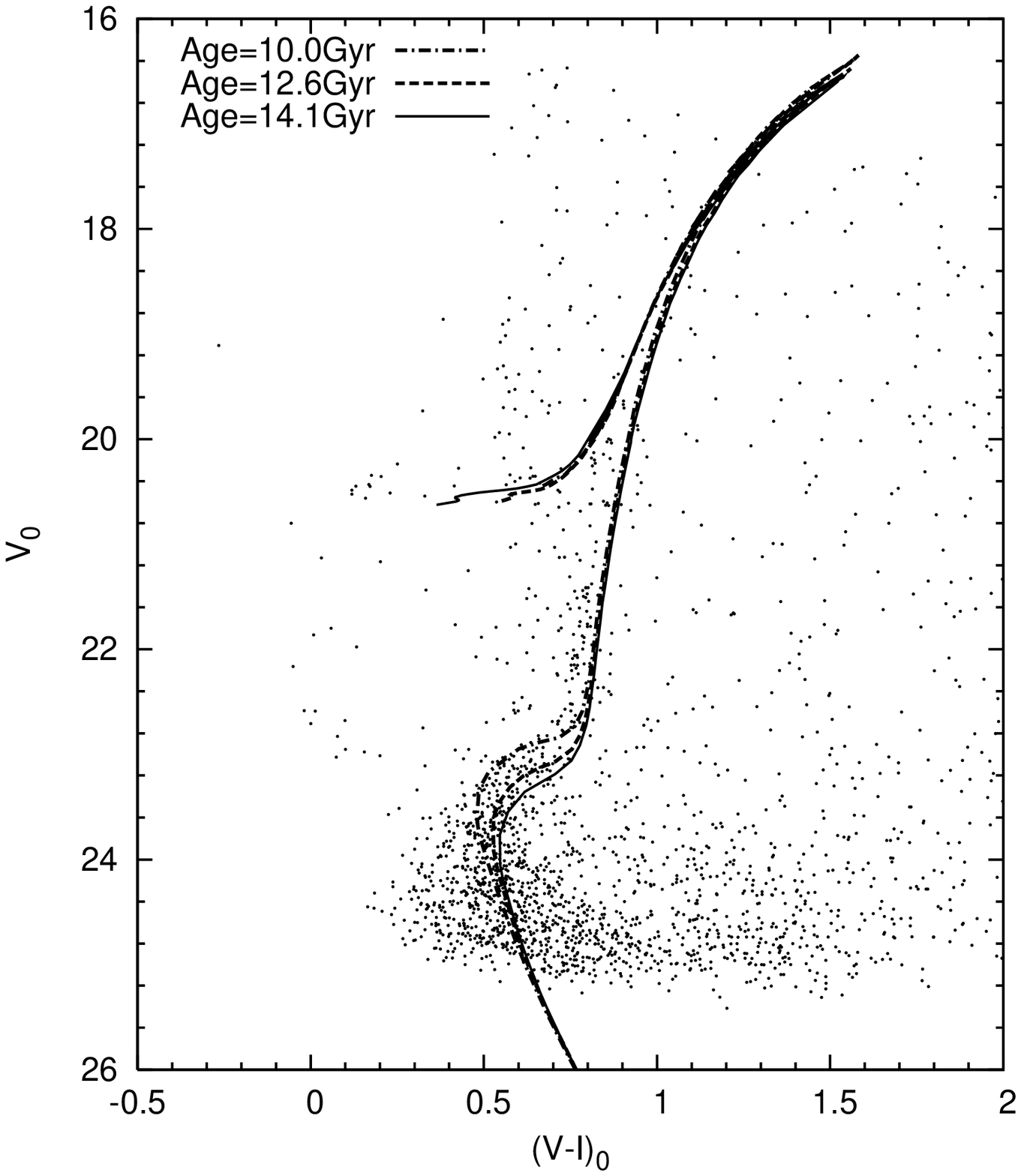}
}
\caption{Left: V-band colour-magnitude diagram of the UMa I dSph with the fiducial sequence of Galactic globular clusters (M3, M5, M92). Right: The theoretical isochrones overlaid in the V-band colour-magnitude diagram of the UMa I dSph.  The lines stand for the interpolated Padova isochrones for [Fe/H]=-2.0 and age of 10.0, 12.6, 14.1 Gyr which were sifted according to the distance of the UMa I dSph. }
\label{fig:cmd iso ggc}
\end{figure*}

These CMDs include foreground stars and background objects.  The main comtamination is the Galactic foreground stars visible at (V$-$I)$_0$ $>$ 0.6 mag.  The contamination of the RGB region in the CMD could in principle be high.  However, the relatively high galactic latidude of the UMa I dSph helps in reducing this effect and the MS, HB features that we consider in figure \ref{fig:cmd} consist predominantly of stars belonging to the UMa I dSph. 
In our radial profile fitting in Section 5, we use the TRILEGAL\footnote{http://trilegal.ster.kuleuven.be/cgi-bin/trilegal} Galaxy model code (Girardi et al. \cite{gir05}) to estimate the foreground star contamination.

%__________________________________________________________________

\section{Distance, Age and Metallicity}

The RGB of the UMa I dSph in figure \ref{fig:cmd} is very thin and the tip of RGB can hardly be identified, so that the RGB tip of the UMa I dSph cannot be used as a distance indicator.  We therefore have estimated the distance to the UMa I dSph using the V-band magnitude of the HB stars.  
We estimate the HB magnitude to $V_{0, HB}=20.45\pm0.02$.  
Since the UMa I dSph seems to have stellar population like Galactic metal-poor globular clusters, we assume theoretical value of absolute magnitude $M_{V,HB}=0.515_{-0.03}^{+0.08}$, corresponding to [Fe/H]$=-2.0_{-0.5}^{+0.5}$ (Catelan et al. \cite{cat04}), which give the distance modulus (distance) as $(m-M)_{0}=19.93\pm0.1$ ($96.8\pm4$ kpc).  

In the left panel of figure \ref{fig:cmd iso ggc}, we compare the V-(V$-$I) diagram with the fiducial sequences of metal-poor Galactic globular clusters M3, M5 and M92, taken from Johnson et al. (\cite{joh98}); the metallicities of these clusters are [Fe/H]=$-1.57, -1.99, -2.24$, respectively (Harris \cite{har96}).  
We adopt the distance moduli to (m$-$M)$_0$ = 14.99, 14.46, 14.67, and reddening corrections to E(B-V) = 0.01, 0.03, 0.02 for M3, M5 and M92, respectively (Harris \cite{har96}).  The mean RGB and MSTO of the UMa I dSph are similar to those of M5 and M92, suggesting that  the metallicity of the UMa I dSph is [Fe/H] $\simeq$ $-$2 and the age is as old as M5 and M92. 

Our metallicity estimate is consistent with previous photometric and spectroscopic studies (Willman et al. \cite{wil05}; Martin et al. \cite{mar07}), and consistent with the metallicity we assumed for estimating the distance ([Fe/H]$=-2.0_{-0.5}^{+0.5}$).  

By overlaying theoretical isochrones on the deep CMD, we estimate the age of stellar population in the UMa I dSph.  The isochrones in the right panel of figure \ref{fig:cmd iso ggc} are interpolated Padova isochrones corresponding to metallicity [Fe/H]=$-$2.0 and age of 10.0, 12.6, 14.1 Gyr (Girardi et al. \cite{gir02}).  
These isochrones are made by the interpolation with the Padova isochrones of metallicity Z=0.0001 and Z=0.0004.  The right panel of figure \ref{fig:cmd iso ggc} shows that the fiducial sequence of MSTO of the UMa I dSph is best reproduced by the isochrone of 12.6 Gyr, consistent with the age roughly estimated by the comparison with Galactic old metal-poor globular clusters in the left panel.  
Consequently, most of stars in the UMa I dSph are estimated to be formed at least 12 Gyrs ago.

The UMa I dSph has old metal-poor stellar population with the narrow dispersion of age and metallicity.  This characteristic of the stellar population of the UMa I dSph seems to be different from ``classical" dSphs, which show evidences of multiple stellar populations of various age and metallicity (e.g. Shetrone et al. \cite{she01}; Ikuta \& Arimoto \cite{iku02}; Tolstoy et al. \cite{tol04}). The UMa I dSph is well-described by a single stellar population, as a typical globular cluster.

%__________________________________________________________________

\section{Structural properties of the UMa I dSph}

\begin{figure}
  \resizebox{\hsize}{!}{\includegraphics{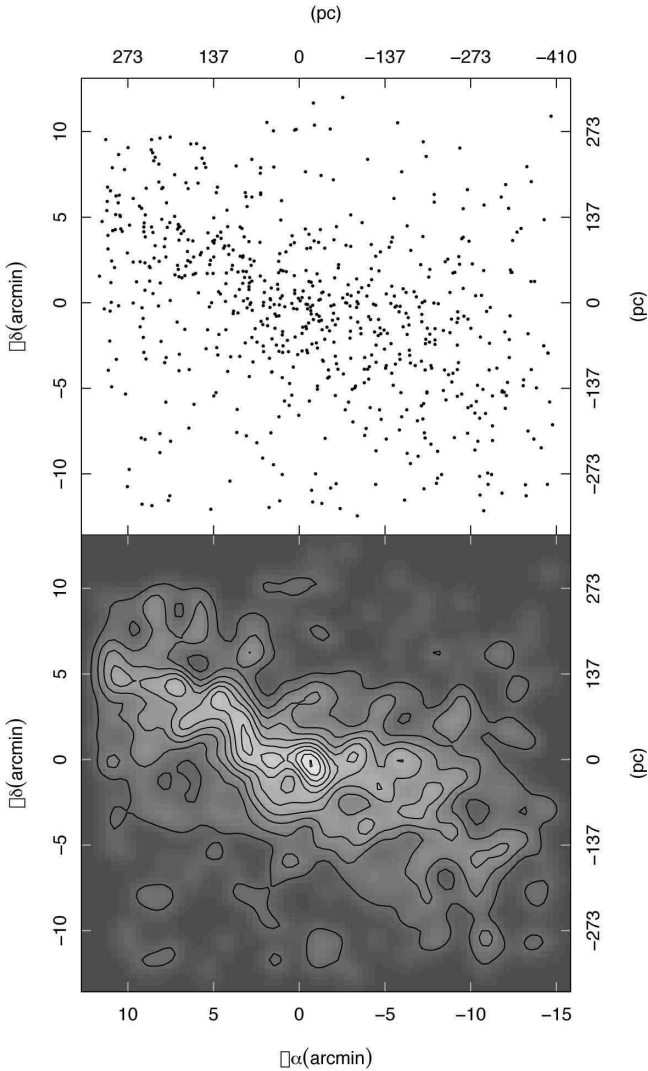}}
  \caption{Upper: The spatial distribution of the stars selected by the dashed lines in figure \ref{fig:cmd}, which coverd the central 730 pc $\times$ 670 pc of the UMa I dSph.  Lower: The isodensity contour of the CMD-selected sources.  The plotted contour levels are 1-10 $\sigma$ above the background level. }
  \label{fig:contour}
\end{figure}

Even though its stellar population closely resembles that of Galactic old metal-poor globular clusters, the spatial distribution of the UMa I dSph shows different properties in morphology and size.  Figure \ref{fig:contour} shows the spatial distribution and the density contour map of stellar sources, which were selected as candidate stars of MS, RGB, HB and BS from the CMD shown in figure \ref{fig:cmd} using the selection boxes marked.  These stars are binned in 30\arcsec$\times$30\arcsec boxes  and smoothed by the Gaussian kernel of bandwidth=0.6\arcmin to draw the lower panel of figure \ref{fig:contour}.  The plotted contour levels are 1-10 $\sigma$ above the background level.  The isodensity contours  covering the central 730 pc $\times$ 670 pc of the UMa I dSph, have quite irregular shapes, stretched toward the upper left and lower right on figure \ref{fig:contour}.  It corresponds to the tangential direction toward the Galactic centre.  

We also compare the spatial distributions of subgiant stars (23.0 $<$ V $<$ 23.4) and faint MS stars (24.0 $<$ V $<$ 24.4) to study the mass segregation in the UMa I dSph,  but there is no difference in both stars.  

\begin{figure}
  \resizebox{\hsize}{!}{\includegraphics[angle=-90]{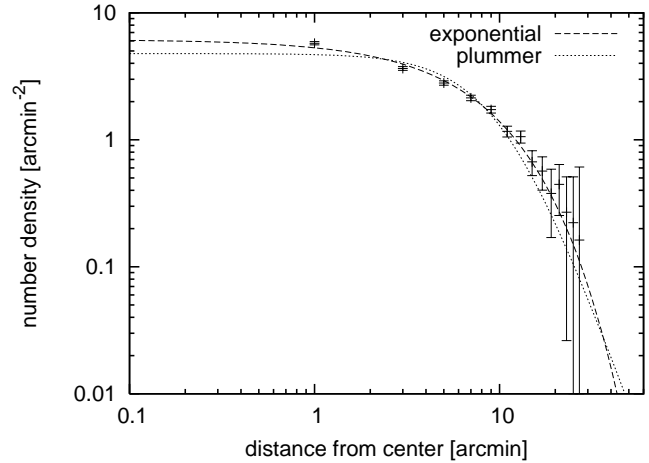}}
  \caption{Raidal profile derived by calculating the average number density within elliptical annuli.  The effect of foreground stars is corrected by using Galactic model (Girardi et al. \cite{gir05}).  Best-fitting exponential profile with the effective radius r$_e$=6.73\arcmin$\pm$0.30\arcmin and Plummer profile with the Plummer radius b=10.42\arcmin$\pm$0.76\arcmin have been overlaid as dashed line and doted line, respectively. }
  \label{fig:radial}
\end{figure}

To calculate the structural parameters, we use the candidate members in figure \ref{fig:cmd} to derive the centroid from the density-weighted first moment of their spatial distribution, and average ellipticity and position angle using the three density-weighted second moments (e.g. Stobie \cite{sto80}). The radial profile shown in figure \ref{fig:radial} is derived by calculating the average number density within elliptical annuli.  The number density of foreground stars in the direction of the UMa I dSph is estimated as 0.05 arcmin$^{-2}$ by using TRILEGAL (Girardi et al. \cite{gir05}).  
We fit the radial profile with standard exponential and Plummer models, and estimate the half-light radius (r$_{1/2}$) of both models as 11.30\arcmin $\pm$ 0.50\arcmin and 10.42\arcmin $\pm$ 0.76\arcmin, respectively.  Hereafter the radius is the elliptical radius $r = [ x^2 + y^2/(1 - e)^2]^{1/2} $, where $e$ is the ellipticity of the galaxy, $x$ and $y$ is the coordinate aligned with the major and minor axis, respectively.  The estimated half-light radius is larger than previous estimation of r$_{1/2}\sim$7.75\arcmin (Willman et al. \cite{wil05}), probably thanks to our much deeper images.  The field level of Willman I (Willman et al. \cite{wil06}) which lies within few degree from the UMa I is within the error bars of field level in our profile.  

The half-light radius of the UMa I dSph corresponds to 300 pc, which is an order of magnitude larger than that of the largest Galactic globular clusters.  The total luminosity of the candidate members is used to estimate the lower limit of the central surface brightness to $\mu_{0,V}$ = 29.5 $\pm$ 0.2 [mag/arcsec$^2$].  These best-fitting structural parameters are listed in table \ref{tbl:1}. 

The structural parameters allow us to constrain the dark matter content of the UMa I dSph, assuming that it is in virial equilibrium. The mass-to-light ratio of a simple stellar system with symmetric distribution can be estimated as M/L=$9\eta\sigma^{2}/2\pi G r_{hb} \Sigma_{0}$, where $\eta$ is a dimensionless parameter dependent on the luminosity distribution, $\sigma$ is a central velocity dispersion, $r_{hb}$ is the half-light radius, and $\Sigma_{0}$ is the central surface brightness (Richstone \& Tremaine \cite{ric86}).  Using the velocity dispersion values $\sigma=9.3[km/s]$ (Kleyna et al. \cite{kle05}) and $\sigma=11.9[km/s]$ (Martin et al. \cite{mar07}), the mass-to-V band light ratio of the UMa I dSph is $M/L\sim1500$ and $2500 M_{\sun}/L_{\sun}$, respectively.  This estimate implies that the UMa I dSph is a highly dark matter dominated galaxy.  However, the shape of the spatial density contours suggests that the UMa I dSph is undergoing tidal disruption.  If this is the case, the galaxy may not be in virial equilibrium.

\begin{table}
\centering 
\caption{Structural parameters of Ursa Major I dSph}
\label{tbl:1}
\begin{tabular}{lc} 
\hline\hline 
Parameter & Value \\
\hline
Coordinates (J2000) & 10h34m44.4s,   $+51^{\circ}55\arcmin33.9\arcsec$\\
Position angle & 78$^{\circ}$.12\\
Ellipticity & 0.54\\
r$_{1/2}$ (exponential) & 11.30 $\pm$ 0.50 [arcmin]\\
r$_{1/2}$ (Plummer) & 10.42 $\pm$ 0.76 [arcmin]\\
$\mu_{0,V}$ (exponential) & 29.5 $\pm$ 0.2 [mag/arcsec$^2$]\\
$\mu_{0,V}$ (Plummer) & 29.7 $\pm$ 0.4 [mag/arcsec$^2$]\\
(m-M)$_{0}$ & 19.93 $\pm$ 0.1\\
Distance & 96.8 $\pm$ 4 [kpc]\\
\hline
\end{tabular}
\end{table}

%__________________________________________________________________

\section{Discussion and summary}

We have used Subaru/Suprime-Cam to obtain multi-colour imaging of the UMa I dSph which is sensitive enough to derive the stellar population below the MSTO and wide enough to study the spatial distribution of stars in the galaxy.  Our deep and wide photometry allows us to study more accurately than previous study.  The distance modulus of the UMa I is derived as (m-M)$_{0}=19.93\pm0.1$, corresponding to 96.8$\pm$4 kpc.  We identify a number of stars in several different evolutionally phases, MS, RGB, HB and BS, which are characterized as a metal-poor single old population.  The mean metallicity is estimated by the V-I color of RGB stars to [Fe/H]$\sim$-2.0, and the turn-off age reveals that the most of stars in the UMa I dSph are formed at least 12 Gyrs ago.  
 
Most ``classical" Galactic dSphs have a sign of multiple stellar populations - different spatial distributions of red and blue HB stars (e.g. Harbeck et al. \cite{har01}; Tolstoy et al. \cite{tol04}), the [$\alpha$/Fe] ratio of most stars \textbf{so far observed} in the dSphs lower than Galactic halo stars of similar metallicity (Shetrone et al. \cite{she01}, \cite{she03}), which implies that star formation in such dSphs lasted for several Gyrs (Ikuta \& Arimoto \cite{iku02}).  On the other hand, the UMa I dSph seems to have a metal-poor single old population, and some of the other newly discovered dSphs also show a sign of similar stellar population 
(Bootes I: Belokurov et al. \cite{bel06}; 
Coma Berenices: Belokurov et al. \cite{bel07}; 
Hercules: Coleman et al. \cite{col07}) 
and appear to have a different property of stellar population from well-known ``classical" dSphs.
Why these dSphs are so faint and have single old population?  If the SN-driven winds blow the gas earlier in such a low-mass galaxy because of the relatively shallow potential, the stellar population may keep the information of the initial star formation activity of the galaxy.  The abundance patterns of these galaxies allow us to reveal the nature of galaxy evolution. 

Although the CMD of the UMa I dSph shows a single epoch of star formation as a metal-poor globular cluster, its spatial size is typical of Galactic dSphs (r$_{1/2}$=300 [pc]), and the shape  suggests that it is undergoing tidal disruption.  The morphology of UMa I dSph is quite irregular, stretched and distorted.  Similar feature is found in the Hercules dSph which shows extremely flat structure ($e\sim0.65$) (Coleman et al. \cite{col07}). If the tidal force could lead such an elongated shape, these galaxies have a high orbital ellipticities and underwent tidal disruption at the last pericenter passage to the Milky Way. 

With the usual assumptions of symmetry and virial equilibrium, the UMa I dSph seems to have high M/L and would be the dark matter dominated object.  If so, the UMa I dSph is presumed to be a long-lived object with distorted morphology.  It is unclear whether the UMa I dSph have been caught in the act of dissolving into the Galactic halo recently, or maintains such elongated shape with dominated dark matter for a long time.  The high-dispersion spectroscopic data will necessary to illuminate the chemical and  dynamical evolution of the galaxy.

\begin{acknowledgements}
We wish to express our gratitude to the anonymous referee for very helpful suggestions and comments. 
The authors are grateful to T. Kodama, Y. Komiyama, L. Greggio and A. Renzini who gave us helpful comments and suggestions.  We also thank the observatory staff of the Subaru Telescope.  S.O. special thanks to M. Iye for great support and comment.  
This work is supported by a Grant-in-Aid for Science Research (No.19540245) by the Japanese Ministry of Education, Culture, Sports, Science and Technology. 
\end{acknowledgements}

\end{document}